%% file: ifacconf.tex
\documentclass{ifacconf}
\input{commands.tex}
\usepackage{graphicx}      % include this line if your document contains figures
\usepackage{natbib}        % required for bibliography
\begin{document}
\begin{frontmatter}

\title{Fast Adaptive Fault Accommodation in Floating Offshore Wind Turbines via Model-Based Fault Diagnosis and Subspace Predictive Repetitive Control\thanksref{footnoteinfo}}
% \title{On the importance of the fault diagnosis in the fast adaptive fault accommodation for floating offshore wind turbines\thanksref{footnoteinfo}} 
% Title, preferably not more than 10 words.

\thanks[footnoteinfo]{This work was supported by the European Union via a Marie Sklodowska-Curie Action (Project EDOWE, grant 835901).}
%% WOULD EDIT AT LAST
\author[First]{Yichao Liu} 
\author[Second]{Ping Wu}
\author[First]{Riccardo M.G. Ferrari}
\author[First]{Jan-Willem van Wingerden}

\address[First]{Delft University of Technology, The Netherlands (e-mail: \{Y.Liu-17, R.Ferrari, J.W.vanWingerden\}@tudelft.nl)}
\address[Second]{Zhejiang Sci-Tech University, China (e-mail: pingwu@zstu.edu.cn)}

\begin{abstract}                % Abstract of not more than 250 words.
\input{sections/0_abstract}
\end{abstract}
%%%%%%%%%%%%%%%%%%%%%%%%%%%%%%%%%%%%%%%%%%%%
%%%%%%%%%%%%%%%%%%%%%%%%%%%%%%%%%%%%%%%%%%%
\end{frontmatter}
%===============================================================================

\section{Introduction}
\input{sections/1_introduction}
%%%%%%%%%%%%%%%%%%%%%%%%%%%%%%%%%%%%%%%%%%

\section{Description of the 10MW FOWT and of fault scenarios} \label{sec:2}
\input{sections/2_FOWT}

%%%%%%%%%%%%%%%%%%%%%%%%%%%%%%%%%%%%%%%%%%

\section{Fast adaptive fault accommodation} \label{sec:3}
\input{sections/3_FD_and_SPRC}
%%%%%%%%%%%%%%%%%%%%%%%%%%%%%%%%%%%%%%%%%%
\section{Case study} \label{sec:4}
\input{sections/4_simulations}

%%%%%%%%%%%%%%%%%%%%%%%%%%%%%%%%%%%%%%%%%%
\section{Concluding remarks} \label{sec:5}
\input{sections/5_conclusions}

%%%%%%%%%%%%%%%%%%%%%%%%%%%%%%%%%%%%%%%%%%
%\begin{ack}
%This work was supported by the EU via a Marie Sklodowska-Curie Action (Project EDOWE, grant 835901).
%\end{ack}

\bibliography{ifacconf}             % bib file to produce the bibliography
                                                     % with bibtex (preferred)
                                                   
%\begin{thebibliography}{xx}  % you can also add the bibliography by hand

%\bibitem[Able(1956)]{Abl:56}
%B.C. Able.
%\newblock Nucleic acid content of microscope.
%\newblock \emph{Nature}, 135:\penalty0 7--9, 1956.

%\bibitem[Able et~al.(1954)Able, Tagg, and Rush]{AbTaRu:54}
%B.C. Able, R.A. Tagg, and M.~Rush.
%\newblock Enzyme-catalyzed cellular transanimations.
%\newblock In A.F. Round, editor, \emph{Advances in Enzymology}, volume~2, pages
%  125--247. Academic Press, New York, 3rd edition, 1954.

%\bibitem[Keohane(1958)]{Keo:58}
%R.~Keohane.
%\newblock \emph{Power and Interdependence: World Politics in Transitions}.
%\newblock Little, Brown \& Co., Boston, 1958.

%\bibitem[Powers(1985)]{Pow:85}
%T.~Powers.
%\newblock Is there a way out?
%\newblock \emph{Harpers}, pages 35--47, June 1985.

%\bibitem[Soukhanov(1992)]{Heritage:92}
%A.~H. Soukhanov, editor.
%\newblock \emph{{The American Heritage. Dictionary of the American Language}}.
%\newblock Houghton Mifflin Company, 1992.

%\end{thebibliography}

%\appendix
%\section{A summary of Latin grammar}    % Each appendix must have a short title.
%\section{Some Latin vocabulary}              % Sections and subsections are supported  
                                                                         % in the appendices.
\end{document}

%% file: commands.tex
\usepackage{graphicx,keyval,trig,url} %,biblatex}
\usepackage[T1]{fontenc}     %LaTeX Font Warning: Font shape `OMS/cmtt/m/n' undefined 
\usepackage[utf8]{inputenc}	%apostrophes appear normally
\usepackage{graphicx}

\let\theoremstyle\relax

\usepackage{amsthm}
\usepackage{amssymb}
\usepackage{amsmath}
\usepackage{mathtools}
\usepackage[mathscr]{eucal}
\usepackage{lipsum}
\usepackage{cuted} % environment changes the formatting from two-column to one-column to better accommodate very long equations
\usepackage{blindtext}
\usepackage{color}
\usepackage{dsfont}

\usepackage{epstopdf}
\usepackage{bm}   %turn the letter into bold in the math mode

\usepackage{color}
\usepackage{xcolor}

\newtheorem{Rem}{Remark}

\newcommand{\R}{\mathbb{R}}

%% file: sections/0_abstract.tex
%% Riccardo's revision
As Floating Offshore Wind Turbines (FOWTs) operate in deep waters and are subjected to stressful wind and wave induced loads, they are more prone than onshore counterparts to experience faults and failure.
%Recently, an adaptive Individual Pitch Control (IPC) technique based on Subspace Predictive Repetitive Control (SPRC) was proposed to accommodate abrupt Pitch Actuator Stuck (PAS) faults.
%Being an adaptive approach, it allowed to compensate a PAS and continue with power generation during faulty conditions.
%A disadvantage of this approach is that it may be slow to adapt after a PAS occurred and thus perform unsatisfyingly.
In particular, the pitch system may experience Pitch Actuator Stuck (PAS) type of faults, which will result in a complete loss of control authority.
In this paper, a novel fast and adaptive solution is developed by integrating a model-based Fault Diagnosis (FD) scheme and the Subspace Predictive Repetitive Control (SPRC).
The FD role is to quickly detect and isolate the failed pitch actuator. Based on the fault isolation results, a pre-tuned adaptive SPRC is switched online in place of the existing one, whose initial values of the parameters has been tuned offline to match the specific faulty case. %thus reducing the online adaptation time.
%In detail, the initial values of the parameters of SPRC has been tuned offline to match the specific failed pitch actuator. The failed pitch actuator is detected and isolated through model-based FD scheme. Based on the fault isolation results, the pre-tuned SPRC is switched online in place of the existing one. 
After that, SPRC employs subspace identification to continuously identify a linear model of the wind turbine over a moving time window, and thereby formulate an adaptive control law to alleviate the PAS-induced loads.
%The effectiveness and benefits of the developed architecture are illustrated using a simulated 10MW FOWT benchmark.
Results show that the developed architecture allows to achieve a considerable reduction of the PAS-induced blade loads. More importantly, the time needed to reduce the PAS-induced loads are significantly shortened, thus avoiding further damage to other components during the adaption time and allowing continued power generation.

%% file: sections/1_introduction.tex
Over the past decade, offshore wind energy has been playing an increasingly important role in the international wind energy mix \citep{GWEC_2019}, being capable of harvesting  deep-water (depth $>60$ m) wind resources.
%
%. With the depletion of coastal resources, geographic limits and environmental constraints, the deployment of the offshore wind harvesting tends to reach deep waters (depth $>60m$) to access abundant offshore wind resources. Floating offshore wind turbines (FOWTs) might be the only feasible solution for harvesting the deep-water wind resources. 
%The first commercial floating wind power project in the world, Hywind Scotland pilot park (\cite{Skaare_2015}), has been on operation since the year of 2017. 
%
However, FOWTs are subjected to continuous and extreme aerodynamic and hydrodynamic loads due to wind and waves, which can lead to unexpected mechanical and electric faults \citep{Carroll_2016}.
%when operating the harsh deep waters with limited access for maintenance (\cite{Liu_2019}).
Particularly, the pitch actuators system, which is critical in optimizing power generation and minimizing structural loads,
% stalling and aerodynamic braking,
account for the biggest proportion (more than 21\%) of the overall failure rate for offshore wind turbines \citep{Jiang_2014}. Consequently, the reliability, safety and resilience of the pitch systems have received increasing attention. 

During operational conditions, the pitch system may experience \emph{severe} faults, such as abrupt Pitch Actuator Stuck (PAS) ones, which may lead to a complete loss of control authority, as well as \emph{non-severe} ones, such as pitch actuator or sensor degradation \citep{Li_2018}. 
Currently, the preferred way to overcome a PAS type of fault currently is via a safe and fast shutdown of the wind turbine \citep{Jiang_2014}. 
However, as PAS faults appear frequently \citep{Ribrant-2006}, such a shutdown solution may lead to high Operation and Maintenance (O\&M) costs due to lost power production and unplanned maintenance.

%These reasons make it urgent to develop a fault accommodation architecture which is able to: 1) accommodate the PAS in a fast adaptive way, 2) prevent further deterioration of the faulty system and 3) make it possible to continue power production until the next maintenance is planned.

%Recently, Subspace Predictive Repetitive Control (SPRC) (\cite{Navalkar_2014}) has been proposed to design an adaptive Individual Pitch Controller (IPC) for wind turbines. SPRC uses subspace identification and data captured over a moving time window to continuously identify a linear model of the wind turbine and design a blade load-limiting control law. By this way it can adapt to slowly changing dynamics.

%However, a disadvantage of that approach is that it can take a long time, dependent on the length of the data time window, to adapt after the occurance of an abrupt fault such a PAS. During such time, improper control actions may be computed which may induce further deterioration of the turbine system.

To approach the dearth of the fault-tolerant control (FTC) for PAS faults, a novel, fast adaptive FTC solution for FOWTs is proposed in this paper, which will reduce blade loads in nominal healthy conditions and rapidly accommodate PAS faults. 
To reach the goal, an \emph{integrated model-based Fault Diagnosis (FD) and Subspace Predictive Repetitive Control (SPRC)} scheme are introduced.
The FD role is responsible for detecting and isolating which pitch actuator failed.
Based on this, the SPRC-based IPC parameters will be switched online to initialize values that were pre-tuned for the specific faulty conditions, by using offline simulations. After the initialization, SPRC utilizes subspace identification and data captured over a moving time window to continuously identify a linear model of the wind turbine and design an adaptive blade load-limiting control law.
%This solution will be shown to speed up the convergence rate of the SPRC-based controller, as illustrated via a numerical study involving a 10MW FOWT benchmark developed by the Technical University of Denmark (DTU) and Stuttgart Wind Energy (SWE) institute \cite{Fontanella_2018a, Fontanella_2018b}.
The effectiveness and benefits of the proposed fast adaptive fault accommodation architecture will be illustrated via a case study involving a 10MW FOWT model \citep{Fontanella_2018a}.

The remainder of the paper is organized as follows. Section \ref{sec:2} presents the 10MW FOWT model and the simulation environment. In section \ref{sec:3}, the SPRC-based IPC and the FD scheme are detailed. Next, a comparison study utilizing a high-fidelity simulator is implemented in Section \ref{sec:4}. Section \ref{sec:5} draws conclusions.

\vspace{-0.15cm}

%% file: sections/2_FOWT.tex
The FOWT model used to demonstrate the benefits of the proposed architecture, is based on the DTU 10MW three-bladed variable speed reference wind turbine and the Triple-Spar floating platform \citep{Fontanella_2018a}.
\begin{figure}
\centering 
\includegraphics[width=1\columnwidth]{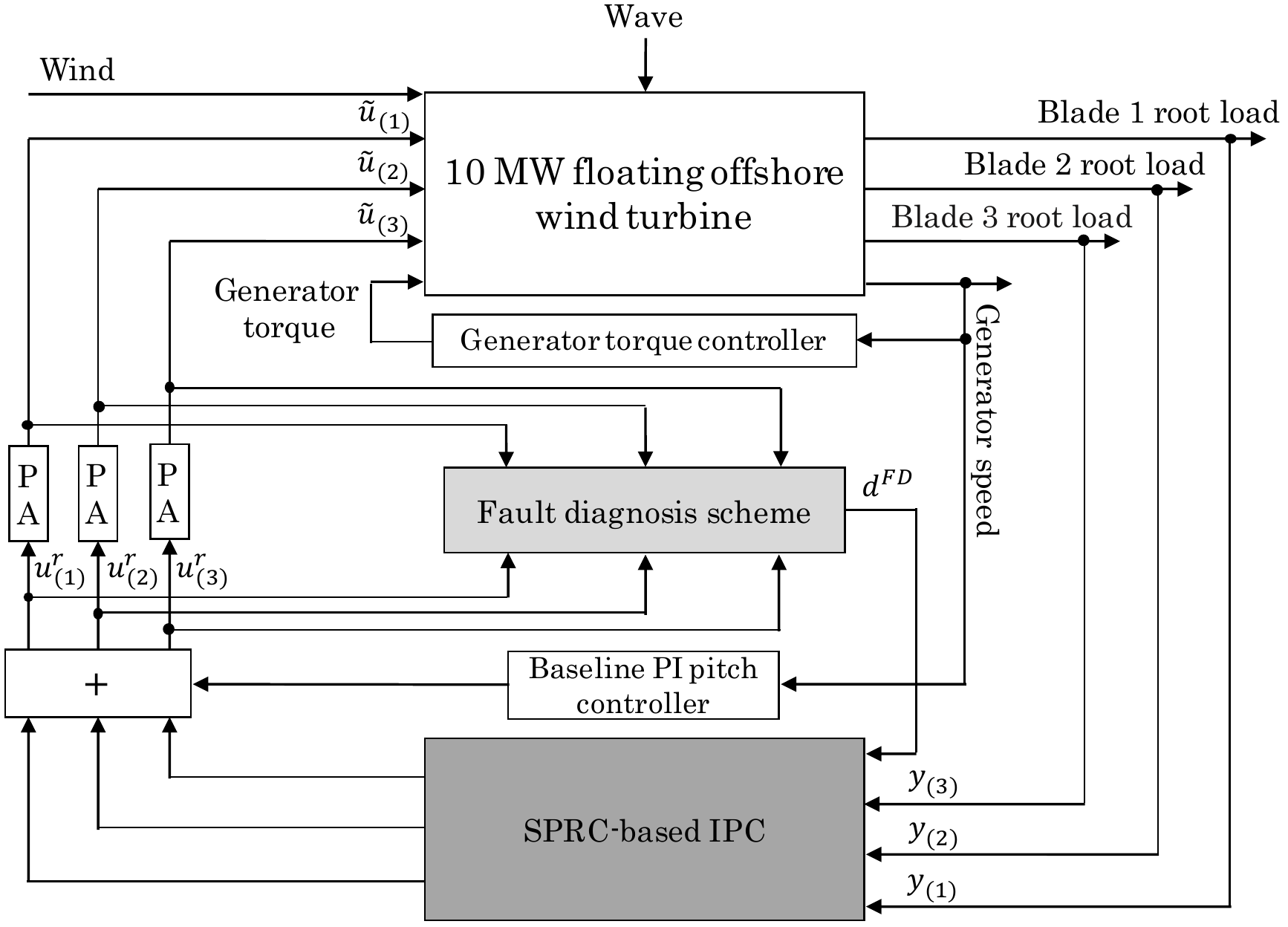}
\vspace{-0.3cm}
\caption{Block diagram of the proposed architecture including FD scheme and SPRC-based IPC, baseline controller and 10MW FOWT. (PA: pitch actuator)}
\label{Pic_block} %
\end{figure}
%%%%%%%%%% end %%%%%%%%%%%%%%%%%%%%

Fig.~\ref{Pic_block} shows the block diagram of the proposed fault tolerant control architecture, which includes a model-based FD block, the switching SPRC-based IPC, a baseline Collective Pitch Controller (CPC, \citealt{Jonkman-2005}) and the 10MW FOWT.
Regarding the simulation environment, the aero-hydro-structural dynamic part of the 10MW FOWT is simulated in the widely-used Fatigue, Aerodynamics, Structures, and Turbulence (FAST) numerical package \citep{Jonkman-2005}, while the baseline wind turbine control part, SPRC-based IPC and FD scheme are implemented in \emph{MathWorks Simulink}. 
In particular, the pitch control utilizes the SPRC-based IPC while the FD scheme uses a model-based approach, both will be introduced in Section \ref{sec:3}, for detection and identification of PAS type of faults. It produces a fault decision $d^{\textrm{FD}} \in \{0,\,1,\,2,\,3 \}$ where $d^{\textrm{FD}}=0$ indicates no fault, and $d^{\textrm{FD}} = l \neq 0$ indicates the $l$--th pitch actuator is faulty. The decision $d^{\textrm{FD}}$ is fed to the SPRC-based IPC block where it is used to switch the controller parameters to offline-learned values as described in the rest of the paper.

The aero-hydro-structural dynamics of the FOWT can be described by the following nonlinear discrete-time system
\begin{align}
\begin{cases}
 x_{k+1} \!\!\! &= A^0x_k+\rho(x_k,\tilde{u}_k) +\eta^{x}(x_k,\tilde{u}_k,k)  \\
 \tilde{u}_k \!\!\! &= u_k +\beta(k-k_0)\phi(u_k,\vartheta) \\
 y_k \!\!\! &= C^0x_k+\eta^{y}(x_k,\tilde{u}_k,k)
\end{cases} \, ,
\label{eq:FOWT_DYNAMICS}
\end{align}
where $k=0,1,\dots$ is the discrete time index and $x\in\R^n$, $\tilde{u}\in\R^q$, $y\in\R^q$ with $q=3$ denote the FOWT state, the control input and the measurement output vectors, respectively. The matrix $A^0\in\R^{n\times{n}}$ and the vector field $\rho:\R^n\times\R^q\mapsto\R^n$ describe the nominal linear and nonlinear parts of the FOWT healthy nominal dynamics while $C^0\in\R^{l\times{n}}$ is the nominal output matrix. The unavoidable modelling uncertainties and periodic disturbances caused by wind loading as well as measurement noise are characterized by the unknown but bounded functions $\eta^{x}:\R^n\times\R^q\times\R\mapsto\R^n$ and $\eta^{y}:\R^n\times\R^q\times\R\mapsto\R^q$. The output $y$ contains the measurements of the three blades root load.

In order to account for the possible effect of a PAS fault, the term $\tilde{u}$ is used to denote the actual physical value of the three blades pitch angles, while $u$ represents the value that would have been produced by a healthy actuator. The two variables are related by the term $\beta(k-k_0)\phi(u_k,\vartheta)$, where $\beta$ is the discrete time unit step, $k_0$ is the unknown index of the fault occurrence time and
\begin{equation}
\begin{gathered}
\phi_k=(-u_k+\vartheta)e^f \, ,
\end{gathered}
\label{eq:PAS}
\end{equation}
is the PAS fault function. The unit vector $e^f\in\R^r$ has a single 1 in its $f$--th position, with $f$ being the index of the stuck actuator while the stuck-at values are contained in $\vartheta\in\R^r$. Finally, the nominal healthy angle $u$ is assumed to depend on the reference value $u^r$ provided by the pitch control system via a second order transfer function
\[
u = \frac{bs+1}{a^2 s^2+bs+1} u^r \,,
\]
with $a=1/\omega^{ac}$ and $b=2\beta^{ac}/\omega^{ac}$, in which angular frequency $\omega^{ac}=6.28\text{rad/s}$ and damping $\beta^{ac}=0.7$.
%
% Furthermore, the term $\beta_{k-k_0}\phi^x(u_k,\vartheta^x)$ represents the changes of the dynamics of the state equation, due to the occurrence of the PAS type of faults at the faulty time index $k_0$. The PAS is modelled by the following equation,
% \begin{equation}
% \begin{gathered}
% \phi^{x}=(-u_{(f)}+\vartheta^{x}_{(f)})e^f \, ,
% \end{gathered}
% \label{eq:PAS}
% \end{equation}
% where $\vartheta^{x}_{(f)}$ is the value of pitch angle of the $f^{\text{th}}$ blade induced by the stuck actuator while $e^f$ denotes an all zeroes column vector of suitable size with a single 1 in its $f^{\text{th}}$ position ($f=1,2,3$).
%In this paper, only one blade is assumed to be stuck at each time. 
\vspace{-0.15cm}

%% file: sections/3_FD_and_SPRC.tex
The theoretical framework behind the proposed architecture %, i.e. integrated SPRC-based IPC and FD scheme, 
will be elaborated in detail in this section. The key points are summarized in the following two steps: 
\emph{Step} 1: An online model-based FD scheme is developed for FOWTs to detect and isolate PAS faults.
%\emph{Step} 2: Critical parameters of the SPRC-based IPC for faulty conditions are offline collected via several numerical training processes.
% \emph{Step} 2: Once a PAS type of fault is diagnosed by the FD scheme, a SPRC-based IPC is based on the FD results.
\emph{Step} 2: A SPRC-based IPC is reconfigured to accommodate PAS faults based on the FD results.

%%%%%%% Fig. %%%%%%%%%%%%%%%%
\begin{figure}
\centering 
\includegraphics[width=0.9\columnwidth]{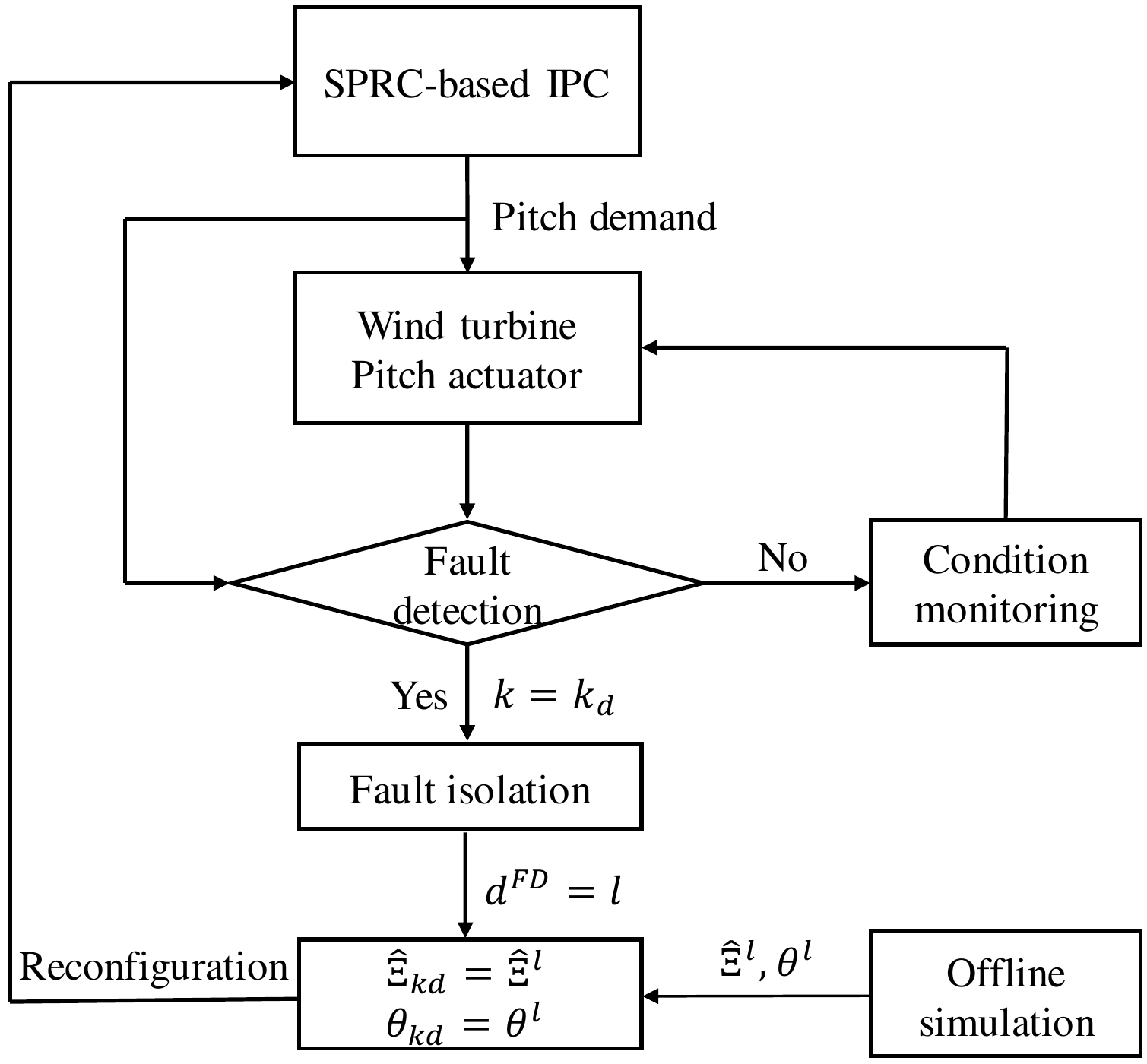}
\vspace{-0.3cm}
\caption{A flowchart showing the main steps and signals involved in the proposed integrated FD and SPRC architecture.}
\label{Pic_algorithm} %
\end{figure}
%%%%% end %%%%%%%%%%%%%%%%%%%

% \vspace{5mm}
% \begin{Rem}\label{rem:fault information}
% The fault information provided for the SPRC-based IPC includes two portions, which is defined here,
% \begin{enumerate}
% \item FD results, i.e. $d^{FD}$, from the online FD scheme, i.e. fault type, detection time, location, as depicted in Fig. \ref{Pic_block}.
% \item Pre-tuned parameters from the offline simulations, for initializing the SPRC algorithm in faulty conditions.
% \end{enumerate}
% \end{Rem}

\subsection{Model-based fault diagnosis for pitch actuator faults}\label{sec:3.1}
The online FD scheme in step 1 is now introduced. Following \cite{Ferrari-2008},
a bank of $N = 3$ Fault Detection and Isolation Estimators (FDIEs), one for each actuator, are designed to yield each an output estimate $\hat{u}^{l}_{k}\in\R$ and a residual signal
\begin{equation}
\begin{gathered}
r^{l}_{k} \triangleq \tilde{u}_{k,(l)}-\hat{u}^{l}_{k} \, ,
\end{gathered}
\label{eq:r}
\end{equation}
where $\tilde{u}_{k,(l)}$ denotes the $l$--th component of the vector $\tilde{u}_{k}\in \R^3$.
The healthy hypothesis would be rejected if the absolute value of at least one residual $r^{l}_{k}$ crosses a suitable time-varying threshold $\bar{r}^{l}_{k}$ at detection time $k_d$.

%\vspace{5mm}
\begin{Rem}\label{rem:extension}
For the sake of simplicity, here only one PAS type of fault is considered in each study case. This means that the $l$--th faulty actuator can be isolated and $d^{\textrm{}FD}$ set to $l$ if the $l$--th residual is the only one crossing its corresponding threshold.
\end{Rem}

The $l$--th FDIE is based on the following discrete-time estimator
%\begin{equation}
%\begin{cases}
%\hat{x}_{k+1,(l)} \!\! &= A_{ac}\hat{x}_{k,(l)}+B_{ac}u_{k,(l)}+L_{ac}(y_{k,(l)}-\hat{y}_{k,(l)}) \\
%\hat{y}_{k,(l)} \!\! &= C_{ac}\hat{x}_{k,(l)}+D_{ac}u_{k,(l)}
%\end{cases} \, ,
%\label{eq:observer}
%\end{equation}
\begin{equation}
\begin{cases}
\hat{x}^{l}_{k+1} \!\! &= A^l\hat{x}^{l}_{k}+B^lu^{r}_{k}+L^l(\tilde{u}_{k,(l)}-\hat{u}^{l}_{k}) \\
\hat{u}^{l}_{k} \!\! &= C^l\hat{x}^{l}_{k}+D^lu^{r}_{k}
\end{cases} \, ,
\label{eq:observer}
\end{equation}
where $\hat{x}^{l}_{k}\in\mathbb{R}^{K}$ and $\hat{u}^{l}_{k}\in\mathbb{R}^{M}$ represent the predicted state vector and output, respectively. $A^l$, $B^l$, $C^l$ and $D^l$, are obtained by implementing a state space realization of the $l$--th actuator dynamics \citep{Fontanella_2018a}. The matrix $L\in\mathbb{R}^{K\times{M}}$ is an estimator gain chosen such that $A^{0,l}\triangleq{A^l-L^lC^l}$ is stable.
A time-varying threshold $\bar{r}^{l}_{k}$, guaranteed to bound the healthy $r^{l}_{k}$, is defined as 
\begin{equation}
\begin{gathered}
\bar{r}^{l}_{k}\triangleq \sum_{h=0}^{k-1}\alpha\delta^{k-1-h}[\bar{\Delta} \rho_{h}+\bar{\eta}^x_{h}]+\alpha\delta^k\bar{\epsilon}^x_{0}
+\bar{\eta}^y_{k}
%, l=l,...,N
\end{gathered} \, ,
\label{eq:threshold}
\end{equation}
where the two constants $\alpha$ and $\delta$ are calculated as in the paper (\cite{Zhang-2001}), such that ${\|C^l(A^{0,l})^k\|}\leq{\alpha\delta^{k}}\leq {\|C^l\|\| A^{0,l} \|^k}$.
The terms $\bar{\eta}^x > \Vert \eta^x \Vert$ and $\bar{\eta}^y > \Vert \eta^y \Vert$ are known upper bounds on the uncertainties. In addition, $\bar{\epsilon}^x_{0} > \Vert x_0 - \hat{x}^l_0 \Vert$ denotes an upper bound on the initial value of the state estimation error. The difference between the real nonlinear dynamics and the value computed by the FDIE is denoted by $\Delta\rho$, with $\bar \Delta\rho$ being its upper bound
\begin{equation}
\begin{gathered}
\Delta\rho(x_k,\hat{x},u_k)\triangleq \rho(x_k,u_k)-\rho(\hat{x}_k,u_k)\,, \\
\bar{\Delta}\rho(\hat{x}_k,u_k)\triangleq \max_{x\in{R^x}}(\|\Delta\rho(x,\hat{x},u_k)\|)
\end{gathered}
\label{eq:threshold_para}
\end{equation}
%\begin{equation}
%\begin{gathered}
%\bar{\epsilon}^{(x)}_k\triangleq \max_{x\in{R^x}}(\|x-\hat{x}_k\|)
%\end{gathered}
%\label{eq:threshold_para2}
%\end{equation}

%In addition, $\bar{\eta}^{(x)}_h$ is,
%\begin{equation}
%\begin{gathered}
%\bar{\eta}^{(x)}_h\triangleq \max_{x\in{R^x}}(\|x-\hat{x}_k\|)
%\end{gathered}
%\label{eq:threshold_para2}
%\end{equation}
% The designed eq. \eqref{eq:threshold} is able to guarantee that no fault-positive alarms will be issued. 

% Based on \eqref{eq:r} to \eqref{eq:threshold}, FD results $d^{FD}$ including (a) Fault type, namely the specified PAS type of faults in the present study, (b) Fault detection instant $k_d$, defined as the first instant in the case where $|r^l_{k}|>\bar{r}^l_{k}$, (c) Fault location, defined as the $l$--th pitch actuator, will be collected and provided for SPRC-based IPC. 
%Consequently, the pitch actuator fault occurring in the $l$--th pitch actuator can be detected and identified according the eq. \eqref{eq:r} if one of FDIE thresholds is crossed by its residual at some time instants (\cite{Ferrari-2008}).

\subsection{Subspace predictive repetitive control}\label{sec:3.2}
After a PAS type of fault is detected and isolated via the designed FD scheme, the SPRC-based IPC \citep{Navalkar_2014} is reconfigured based on the FD results to accommodate the fault, as summarized in step 2. 
In detail, the dynamics of the SPRC-based IPC of FOWTs under faulty conditions can be described by the following LTI system affected by unknown periodic disturbances \citep{Houtzager_2013}. In prediction form, it is
%\begin{equation}
%\begin{cases}
%\label{eq:LTI}
%x_{k+1} \!\! &= Ax_k+B(u_k+\phi_{k})+Ed_k+Le_k \\
%y_k \!\! &= Cx_k+Fd_k+e_k
%\end{cases} \, ,
%\end{equation}
%where $x_k\in\mathbb{R}^n$, $u_k\in\mathbb{R}^r$ and $y_k\in\mathbb{R}^l$ denote the state, control input and output vectors. In particular, $l=r=3$, $u_k$ includes the blades pitch angles and $y_k$ contains the blade loading (i.e. Out-of-Plane bending moment (MOoP)) at discrete time index $k$. 
%In a predictor form, the eq. (\ref{eq:LTI}) can be rewritten as,
\begin{equation}
\begin{cases}
\label{eq:predictor:f}
x_{k+1} \!\! &= \tilde{A}x_k+B(u_k+\phi_{k})+\tilde{E}d_k+Ly_k \\
y_k \!\! &= Cx_k+Fd_k+e_k
\end{cases} \, ,
\end{equation}
where $d_k$ denotes the periodic component of disturbances on the blades, while $e_k\in\mathbb{R}^l$ is the aperiodic component of the blade loading. 
In addition, $\tilde{A}\triangleq{A-LC}$ and $\tilde{E}\triangleq{E-LF}$, where matrices $A\in\mathbb{R}^{n\times n}$, $B\in\mathbb{R}^{n\times r}$, $C\in\mathbb{R}^{l\times n}$, $L\in\mathbb{R}^{n\times l}$, $E\in\mathbb{R}^{n\times m}$ and $F\in\mathbb{R}^{l\times m}$ represent state transition, input, output, observer, periodic noise input and periodic noise direct feed-through matrices, respectively. During healthy conditions ($0\leq{k}<{k_0}$), it holds $\phi_k=0$.
%Similarly, the LTI system with PAS type of faults in the faulty condition with time instants $k_d\leq{k}$ is formulated as,
The effect of the periodic disturbance on the input-output system could be eliminated by defining a periodic difference operator $\delta$ as,
\begin{align*}
 \delta{d}_k &=d_k-d_{k-P}=0, \\
 \delta{u}_k &= (u_k+\phi_k) - (u_{k-P}+\phi_{k-P}), \\
 \delta{y}_k &=y_k-y_{k-P},
%\label{eq:difference operator}
\end{align*} 
where $P$ denotes the disturbance period. 
During the occurrence of a PAS type of fault, $\delta{u}_{k,(f)}$ for the $f$--th blade is 0, since $\phi_k=\phi_{k-P}$ accordingly.

Based on the definition of $\delta$, eq. \eqref{eq:predictor:f} is formulated as
\begin{equation}
\begin{cases}
\delta{x}_{k+1} \!\! &= \tilde{A}\delta{x}_k+B\delta{u}_k+L\delta{y}_k \\
\delta{y}_k \!\! &= C\delta{x}_k+\delta{e}_k
\end{cases} 
\label{eq:predictor2}
\end{equation}

Considering a given time window of length $p$ in the past, the following stacked vector can be defined, %$\delta{U}^{[P]}_{k}$
\begin{equation}
\delta{U}^{[p]}_{k}=
\left[ \begin{array}{c}
 u_k-u_{k-P}\\
u_{k+1}-u_{k-P+1} \\
\vdots \\
u_{k+p-1}-u_{k+p-P-1}
\end{array} 
\right ]\, ,
\label{eq:stacked u}
\end{equation}
and, similarly, the vector $\delta{Y}^{[p]}_{k}$. 
If $p$ is large enough such that $\tilde{A}^{j}\approx0$ $\forall{j}\geq{p}$ (\cite{Chiuso_2007}), the future state vector $\delta{x}_{k+p}$ can be approximated based on $\delta{U}^{[p]}_{k}$ and $\delta{Y}^{[p]}_{k}$ as
%\begin{equation}
%\delta{x}_{k+p}= \tilde{A}^{p}\delta{x}(k)+
%\left[ \begin{array}{cc}
%K^{[p]}_u & K^{[p]}_y 
%\end{array} 
%\right ]
%\left[ \begin{array}{c}
%\delta{U}^{[p]}_{k} \\
%\delta{Y}^{[p]}_{k} \\
%\end{array} 
%\right ]\, ,
%\label{eq:lifted}
%\end{equation}
\begin{equation}
\delta{x}_{k+p}=
\left[ \begin{array}{cc}
K^{[p]}_u & K^{[p]}_y 
\end{array} 
\right ]
\left[ \begin{array}{c}
\delta{U}^{[p]}_{k} \\
\delta{Y}^{[p]}_{k} \\
\end{array} 
\right ]\, ,
\label{eq:lifted2}
\end{equation}
where $K^{[p]}_u$ and $K^{[p]}_y$ are defined as,
\begin{align*}
&\ K^{[p]}_u=
\left[ \begin{array}{cccc}
\tilde{A}^{p-1}B & \tilde{A}^{p-2}B & \cdots & B 
\end{array} 
\right ]\,,\\ 
&\ K^{[p]}_y=
\left[ \begin{array}{cccc}
\tilde{A}^{p-1}L & \tilde{A}^{p-2}L & \cdots & L
\end{array} 
\right ]\,.
\label{eq:K}
\end{align*}

%It is worth noting that $p$ should be large enough such that $\tilde{A}^{j}\approx0$ if $\forall{j}\geq{p}$ (\cite{Chiuso_2007}). In such a case, $\delta{x}_{k+p}$ is approximately equal to

Combining eq. (\ref{eq:lifted2}) with (\ref{eq:predictor2}), $\delta\hat{y}_k$ can be estimated as
\begin{equation}
\delta{y}_{k+p}=
\underbrace{
\left[ \begin{array}{cc}
CK^{[p]}_u & CK^{[p]}_y 
\end{array} 
\right ]}_{\Xi}
\left[ \begin{array}{c}
\delta{U}^{[p]}_{k} \\
\delta{Y}^{[p]}_{k} \\
\end{array} 
\right ]
+\delta{e_{k+p}}  \,,
\label{eq:lifted3}
\end{equation}
where the Markov matrix $\Xi\in\mathbb{R}^{l\times((r+l)\cdot{p})}$ was introduced.
%It is clear from the eq. (\ref{eq:lifted3}) that the matrix of coefficients $\left[\begin{array}{cc} CK^{[p]}_u & CK^{[p]}_y\end{array}\right]$ contains all the necessary information on the FOWT dynamics and can be approximated by the input $u$ and output $y$. It is then defined as FOWT Markov parameters $\Xi\in\mathbb{R}^{l\times((r+l)\cdot{p})}$ as, 
%\begin{equation}
%\Xi=
%\left[ \begin{array}{cc}
%CK^{[p]}_u & CK^{[p]}_y 
%\end{array} 
%\right ]
%\label{eq:Markov parameters}
%\end{equation}
%
In essence, the aim of the identification is to find an online solution of a Recursive Least-Squares (RLS) optimization problem. 
%By starting from eq. \eqref{eq:lifted3} and using the measurements of $u$ and $y$, the following problem is formulated here,
%\begin{equation}
%\hat{\Xi}_k=\text{arg}\min_{\hat{\Xi}_k}\sum_{k=0}^{\infty}\left\lVert\delta{y}_k-\lambda%\hat{\Xi}_k
%\left[ \begin{array}{c}
%\delta{U}^{[p]}_{k} \\
%\delta{Y}^{[p]}_{k} \\
%\end{array} 
%\right ]
%\right\rVert^2_2 \, ,
%\label{eq:Markov parameters2}
%\end{equation}
%($\lambda$ represents a window function of $10^6$ samples (\cite{Fredrik_2000})). 
In order to achieve adaptive tolerant control for the PAS type of fault, the RLS optimization is decoupled for each blade based on the assumption that the $n$--th blade load is independent from the $m$-th, where $n\neq{m}$. 
Therefore, the subspace identification step for faulty conditions is implemented by the following RLS optimization 
\begin{equation}
\hat{\Xi}_{k,(l)}=\text{arg}\min_{\hat{\Xi}_k}\sum_{k=0}^{\infty}\left\lVert\delta{y}_{k,(l)}-\lambda\hat{\Xi}_{k,(l)}
\left[ \begin{array}{c}
\delta{U}^{[p]}_{k,(l)} \\
\delta{Y}^{[p]}_{k,(l)} \\
\end{array} 
\right ]
\right\rVert^2_2 \, ,
\label{eq:Markov parameters3}
\end{equation}
where $\lambda$ denotes a forgetting factor ($0\ll\lambda\leq{1}$) to attenuate the effect of past data, and adapt to the updated system dynamics online. In this paper, a large value, i.e. $\lambda=0.99999$, was selected to guarantee the robustness of the optimization process \citep{Fredrik_2000}.  
$l=1,2,3$ represents the blade number, while $\hat{\Xi}_{k,(l)}$ is the estimate of independent Markov matrix for each blade. As a consequence, the optimization process in eq. (\ref{eq:Markov parameters3}) is conducted three times at each time instant $k$. Next, the $\hat{\Xi}_{k}$ is synthesized as $\hat{\Xi}_{k}=[\hat{\Xi}_{k,(1)}, \hat{\Xi}_{k,(2)}, \hat{\Xi}_{k,(3)}]^T$.
%\begin{equation}
%\hat{\Xi}_{k}=
%\left[ \begin{array}{c}
%\hat{\Xi}_{k,(1)} \\
%\hat{\Xi}_{k,(2)} \\
%\hat{\Xi}_{k,(3)} \\
%\end{array} 
%\right ]
%\label{eq:Markov parameters4}
%\end{equation}
\iffalse
Consequently, $\hat{\Xi}_k$ contains estimates of the following matrices based on RLS optimization, as
%\begin{equation}
%\hat{\Xi}_k=
%\left[ \begin{array}{cccccccc}
%\widehat{CA^{p-1}B} & \widehat{CA^{p-2}B} & \cdots & \widehat{CB} & 
%\widehat{CA^{p-1}K} & \widehat{CA^{p-2}K} & \cdots & \widehat{CK} 
%\end{array} 
%\right ] \, ,
%\label{eq:estimates of Markov parameter}
%\end{equation}
\begin{multline}
\hat{\Xi}_k=
\biggl[ 
\begin{array}{cccc}
\widehat{C\tilde{A}^{p-1}B} & \widehat{C\tilde{A}^{p-2}B} & \cdots & \widehat{CB}
\end{array}
\\
\begin{array}{cccc}
\widehat{C\tilde{A}^{p-1}K} & \widehat{C\tilde{A}^{p-2}K} & \cdots & \widehat{CK}
\end{array}
\biggr] 
\label{eq:estimates of Markov parameter}
\end{multline}
\fi
Therefore, the FOWT system dynamics are identified online, taking into consideration the faulty conditions due to the occurrence of PAS.
It is noted that the FOWT system should be persistently excited in order to obtain a unique solution of the RLS optimization \citep{Michel_2007}. Then, the RLS problem is solved via a QR algorithm, as introduced in  \cite{Sayed_1998}. Consequently, the estimates of $\hat{\Xi}_k$ are employed to formulate a SPRC law.

The state feedback controller can be formulated with a state-space representation \citep{Navalkar_2014} based on the identified $\hat{\Xi}_k$,
\begin{multline}
\underbrace{
\left[ \begin{array}{c}
\bar{Y}_{j+1}\\
\delta{\theta}_{j+1} \\
\delta{\bar{Y}}_{j+1} \\
\end{array} 
\right ]}_{\bar{K}_{j+1}}
=
\underbrace{
\left[ \begin{array}{ccc}
I_{l\cdot{P}} & \phi^{+}\widehat{\Gamma^{[P]}K^{[P]}_u}\phi & \phi^{+}\widehat{\Gamma^{[P]}K^{[P]}_y}\phi \\
0_{l\cdot{P}} & 0_{r\cdot{P}} & 0_{l\cdot{P}} \\
0_{l\cdot{P}} & \phi^{+}\widehat{\Gamma^{[P]}K^{[P]}_u}\phi & \phi^{+}\widehat{\Gamma^{[P]}K^{[P]}_y}\phi 
\end{array} 
\right]}_{\bar{A}_j}
\\
\underbrace{
\left[ \begin{array}{c}
\bar{Y}_j \\
\delta{\theta}_j \\
\delta{Y}_j
\end{array} 
\right]}_{\bar{K}_j}
+
\underbrace{
\left[ \begin{array}{c}
\phi^{+}\hat{H}^{[P]}\phi \\
I_{r\cdot{P}} \\
\phi^{+}\hat{H}^{[P]}\phi
\end{array} 
\right]}_{\hat{B}_j}
\delta{\theta}_{j+1}
\, ,
\label{eq:state-space form_lower}
\end{multline}
where $j=0,1,2,\cdots$ is the rotation count of the rotor.
$\hat{H}^{[P]}$ and $\Gamma^{[P]}$ are the same matrices defined in the paper (\cite{Navalkar_2014}).
The symbol $+$ represents the Moore-Penrose pseudo-inverse.
$\theta \in \mathbb{R}^{2r}$ denotes the control inputs projected on the basis function $\phi$, analogously to the study (\cite{Wijdeven_2010}),
\begin{equation}
U^{[P]}_k=\phi\cdot\theta_{j}
\, .
\label{eq:control input}
\end{equation}
It is worth noting that $\theta$ is updated at each $P$. The state transition and input matrices are updated at each discrete time instance $k$.
Based on this, the classical optimal state feedback matrix ${K}_{f,j}$ can be synthesised in a Linear Quadratic Regulator (LQR) sense (\cite{Hallouzi_2006}).
%In this paper, only 1P loads are considered.
%Next, a transformation matrix of the basis function projection  (\cite{Wijdeven_2010}) $\Omega\in\mathbb{R}^{(r\cdot{P})\times{2r}}$ is utilized to limit the spectral content of $U_k$ within the desired frequency range and reduce the dimension of the Discrete Algebraic Riccati Equation (DARE). In this paper, only 1P loads are considered.
%In addition, such a projection can reduce the dimension of the Discrete Algebraic Riccati Equation DARE that must be solved, thus reducing the computational cost. 
%In this paper, only 1P loads were considered, based on the following transformation matrix $\phi\in\Re^{(r\cdot{P})\times{2r}}$,
%\begin{equation}
%\phi=
%\underbrace{
%\left[ \begin{array}{cc}
%sin(2\pi/P) & cos(2\pi/P) \\
%sin(4\pi/P) & cos(4\pi/P) \\
%\vdots & \vdots \\
%sin(2\pi) & cos(2\pi)
%\end{array} 
%\right ]}_{U_f} 
%\otimes{I}_r
%\, ,
%\label{eq:with basis function}
%\end{equation}
%where $U_f$ denotes the basis function and the symbol $\otimes$ represents the Kronecker product. In order to take into account the rotor speed variations (that might be induced by occurrence of faults, wind turbulence, changes in the inflow wind speed...), the rotor azimuth $\psi$, which is equal to $2\phi{k}/P$ at time instant $k$, is used instead of the fixed $U_f$. 
Given the state feedback law, the control signal for the frequency of interest, e.g. 1P, is formulated, as
\begin{equation}
\delta\theta_{j+1}=-K_{f,j}\bar{K}_j
\, ,
\label{eq:control input2}
\end{equation}
Considering that $\delta\theta_{j+1}=\theta_{j+1}-\theta_{j}$, the projected output update law $\theta_{j+1}$ can be calculated as,
\begin{equation}
\theta_{j+1}=\sigma\theta_{j}-\beta{K}_{f,j}
\left[ \begin{array}{c}
\bar{Y}_{j} \\
\delta\theta_{j} \\
\delta\bar{Y}_{j}
\end{array} 
\right]
\, ,
\label{eq:control input3}
\end{equation}
where $\sigma\in[0,1]$ and $\beta\in[0,1]$ are tuning parameters related to the convergence rate of the algorithm. $K_{f,j}$ denotes the optimal state feedback gain during the formulation of the state feedback controller. 
%Finally, the adaptive SPRC control law is then generated according to eq. \eqref{eq:control input}, thus leading to load reduction in healthy conditions and to adaptive accommodation of PAS kind of faults.

%In summary, the fast adaptive accommodation can be implemented based on the theoretical framework introduced in section \ref{sec:3}. 
%In \emph{Step} 1, critical matrix and parameter, i.e. $\hat{\Xi}_f$ and $\theta_f$, of the SPRC-based IPC corresponding to the detected PAS type of fault are collected offline.
%Then, the FD scheme developed in section \ref{sec:3.1} is utilized to detect and identify PAS type of faults, as described in \emph{Step} 2. 
%Once the PAS type of fault is successfully detected and identified via the developed FD scheme, necessary information on the detected PAS type of fault will be given to the SPRC-based IPC from FD scheme. 
%Finally, the collected $\hat{\Xi}_f$ and $\theta_f$ corresponding to the PAS type of fault are provided to be the initial conditions for the SPRC-based IPC as elaborated in \emph{Step} 3, in order to reconfigure the controller immediately, and thereby achieve a fast adaptive fault accommodation. 
%In order to achieve an immediate reconfiguration of aforementioned SPRC algorithm, the parameters of SPRC (i.e. $\hat{\Xi}^f$ and $\theta^f$), which essentially represent the dynamics of SPRC-based IPC caused by the PAS type of faults, are pre-tuned from the offline simulations. 
%Provided with the fault information from the online FD scheme and offline tuned parameters, fast adaptive fault accommodation is then realized.
If $l$--th faulty actuator is isolated at $k_d$ and $d^{FD}=l$, the pre-tuned values of SPRC parameters, i.e. $\theta^{l}$ and $\hat{\Xi}^{l}$, are switched online to initialize the controller as, 1) $\theta_{k_d}=\theta^{l}$, 2) $\hat{\Xi}_{k_d}=\hat{\Xi}^{l}$.
%\begin{equation}
%\begin{cases}
%\theta_{k_d}=\theta^{l} \\
%\hat{\Xi}_{k_d}=\hat{\Xi}^{l}
%\, .
%\end{cases} 
%\label{eq:pre_tuned}
%\end{equation}
A flow-chart of the proposed integrated FD and SPRC architecture is presented in Fig. \ref{Pic_algorithm}.

\vspace{-0.4cm}

%% file: sections/4_simulations.tex
The effectiveness and benefits of the developed fast adaptive fault accommodation architecture is verified in this section, via a case study on the 10MW FOWT (Fig. \ref{Pic_block}). 

\subsection{Model configuration}
% As portrayed in Fig. \ref{Pic_block}, the aero-hydro-structural dynamics of the 10MW FOWT were simulated in \emph{FAST v8.16} (\cite{Jonkman-2005}), while the proposed fast adaptive fault accommodation architecture and other controllers were implemented in \emph{Simulink}. 
In total, three Load Cases (LCs), which are characterized by a uniform  wind profile, are considered in the case study. The mean hub-height wind speed $U_{\text{hub}}$ are 12, 16 and 20 m/s respectively. 
In addition, one specific PAS type of fault is chosen for each LC, considering a different pitch angle setting $\vartheta_{(3)}$ equaling to 20$^\circ$, 0$^\circ$, 10$^\circ$ respectively for the stuck blade ($f=3$ in all cases). 
During each LC, totally 1400s are simulated at a fixed discrete time step of $T_s =$ 0.01 s, with a fault occurring at $T_0=$900 s. 
The measured signal is pitch angle for each blade, which was affected by a Gaussian white noise measurement with variance of $1.5^\circ$.

In order to guarantee persistence of excitation in the nominal healthy and faulty conditions for online subspace identification in the SPRC-based IPC, a filtered pseudo-random binary signal with a maximum amplitude of $3^{\circ}$ is superimposed on top of the collective pitch demand of blades. 
Based on the excited FOWT system, the Markov matrix is recursively updated by the RLS algorithm and then used for the generation of the SPRC control law.
%Moreover, the value of the past window $p$ was selected as $21$.
%In order to ensure a successful implementation of the fast adaptive fault accommodation, the matrix $\hat{\Xi}_f$ and parameter $\theta_f$ of faulty conditions are collected for each load case offline. 
%With the aid of the FD scheme and collected $\hat{\Xi}_f$ and $\theta_f$, the fast adaptive fault accommodation for each load case, with different random seed of measurement noise, is simulated for investigations. 

%It is worth noticing that the online subspace identification is implemented until adequate information on the FOWT dynamics is collected. Moreover, the value of past window $p$ was selected as 21, in order to increase the convergence rate of the subspace identification.
\subsection{Fast adaptive fault accommodation}
\begin{figure*}
\centering 
\includegraphics[width=2\columnwidth]{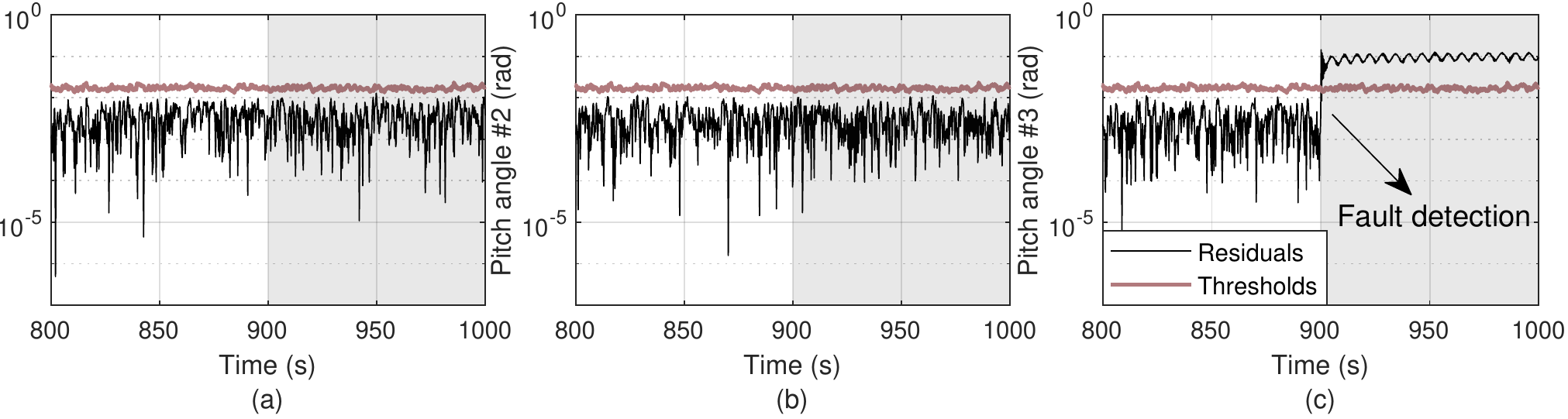}
\vspace{-0.3cm}
\caption{Results of fault diagnosis for PAS type of faults. (a) Pitch angle of blade $\#$1, (b) Pitch angle of blade $\#$2, (c) Pitch angle of blade $\#$3. Time periods of faulty conditions are indicated by a grey background.}
\label{Pic_FD} %
\end{figure*}

%\subsection{Model-based fault diagnosis}
In order to appreciate the performance of the designed model-based FD scheme, detection results for LC3 are shown in Fig. \ref{Pic_FD}.
%which would be the precondition of the subsequent fast adaptive fault accommodation.
Before $T_0 = 900$ s, all the residuals are bounded by their corresponding thresholds, thus verifying the robustness of the threshold. 
After the fault time $T_0$, only the pitch angle residual of blade $\#$3 crosses the corresponding threshold, while others are still bounded by their thresholds (see Fig. \ref{Pic_FD}(a-b)), which imply that the PAS fault is successfully detected and isolated. Hence, the correct fault decision $d^{FD}=l=3, l=f$ is obtained and used for reconfiguring the SPRC-based IPC.
%Other LCs show similar patterns and thereby omitted for brevity. 

%%%%% Fig. FD %%%%%%
%%%%% end %%%%%%%%%%

\begin{figure}
\centering 
\includegraphics[width=1\columnwidth]{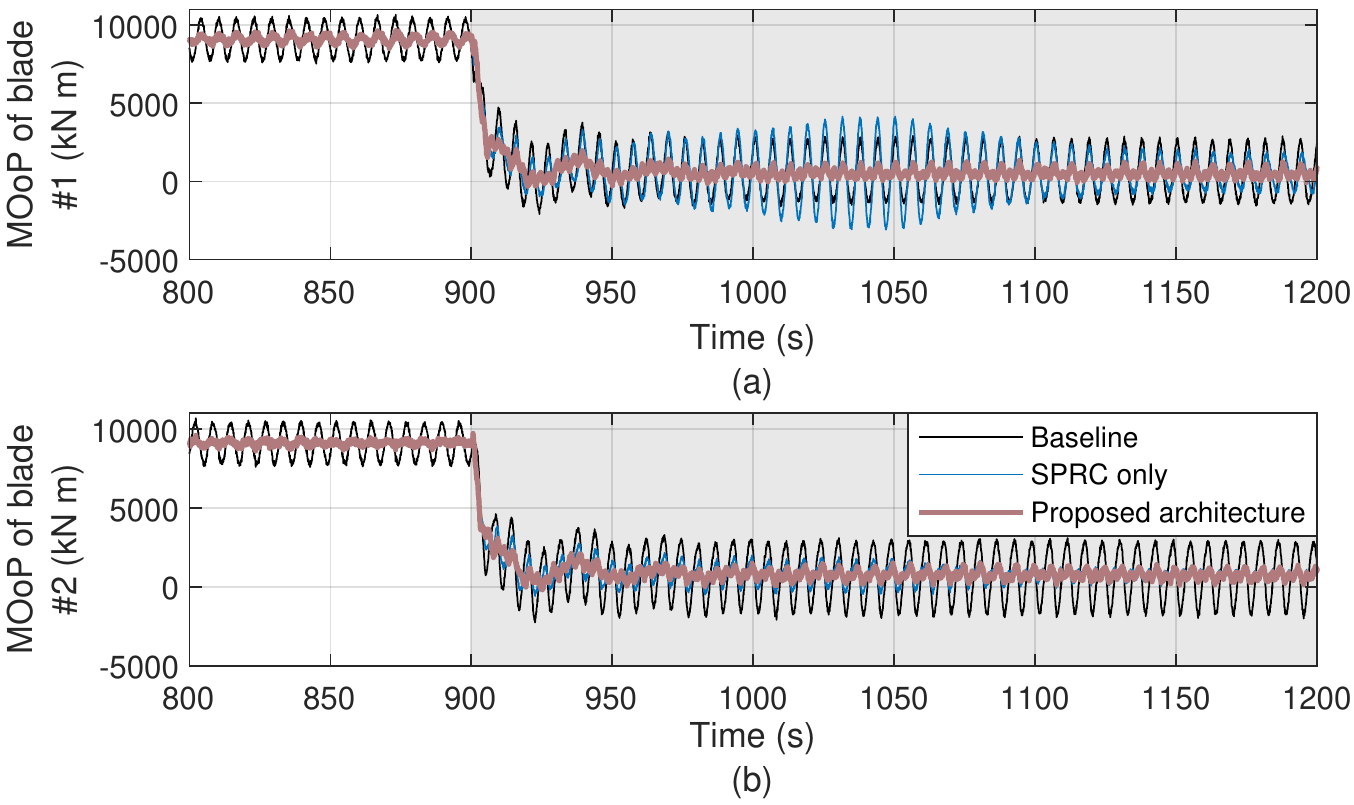}
\vspace{-0.3cm}
\caption{Blade root MOoP. (a) Blade $\#$1, (b) Blade $\#$2. Time periods of faulty conditions are indicated by a grey background. Blade $\#$3 is not shown since it is the faulty blade.}
\label{Pic_MOOP} %
\end{figure}

\begin{figure}
\centering 
\includegraphics[width=1\columnwidth]{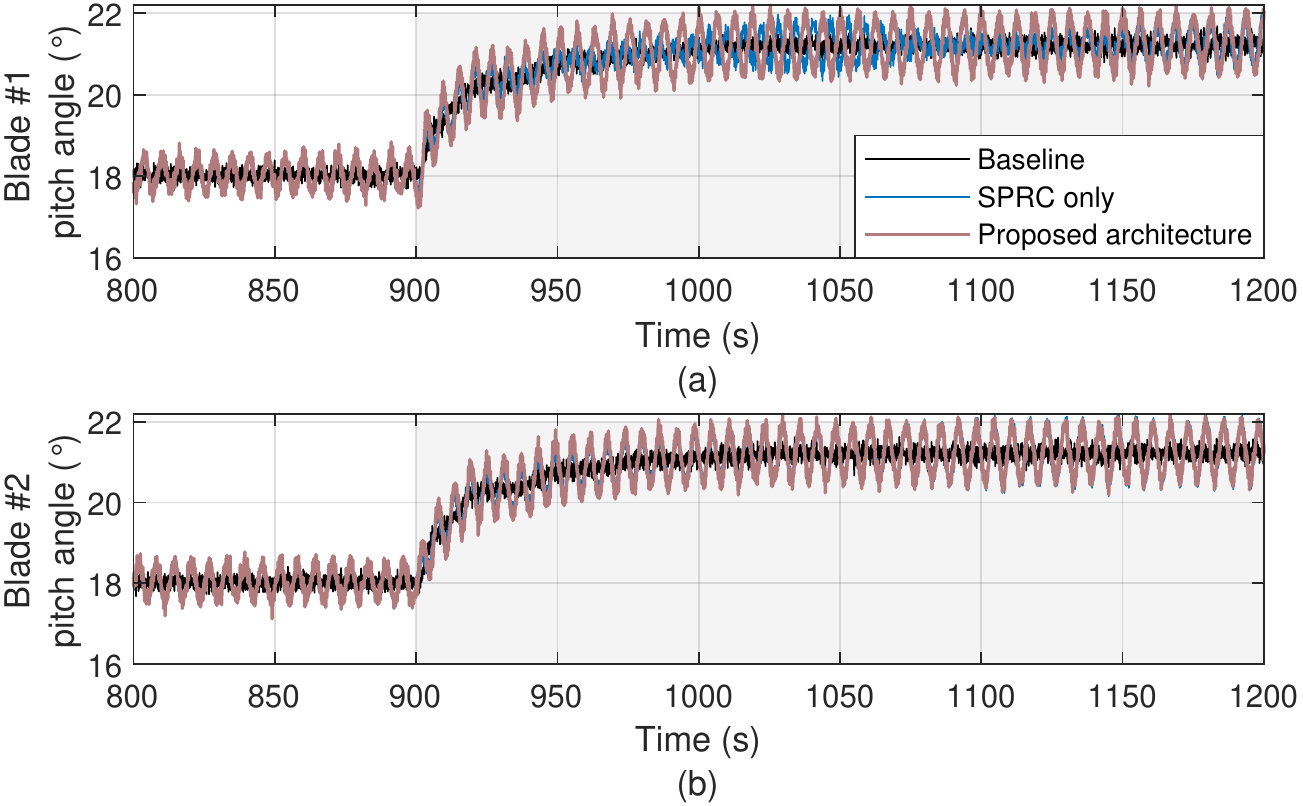}
\vspace{-0.3cm}
\caption{Pitch angles. (a) Blade $\#$1, (b) Blade $\#$2. Time periods of faulty conditions are indicated by a grey background.  Blade $\#$3 is not shown since it is the faulty blade.}
\label{Pic_angle} %
\end{figure}

%\subsection{Fast adaptive fault accommodation}
Based on the FD results, pre-tuned parameters for the SPRC-based IPC are switched online in order to quickly accommodate the fault. Such new initial values, i.e. $\hat{\Xi}^l$ and $\theta^l$, have been tuned offline to match the specific faulty blade, thereby reducing the adaptation time. 
%in order to speed up the convergence of the algorithm, and thereby realize a fast adaptive fault accommodation of PAS fault.  
%Particularly, $\hat{\Xi}^f$ and $\theta^f$ are collected from the offline tuning process.
For the offline tuning purpose, two FAST simulations per each possible fault were run to determine $\hat{\Xi}^l$ and $\theta^l$. In particular the specific PAS type of fault is injected at the beginning of the simulated steady operation of the 10MW FOWT. As a result, time series of $\theta$ in offline simulations, showing the convergence of the SPRC algorithm, are collected.
In order to demonstrate the effectiveness and benefits of the proposed architecture, Figs. \ref{Pic_MOOP}-\ref{Pic_angle} show the comparisons of MOoP and pitch angles between the proposed architecture and other two control logics (i.e. baseline controller and SPRC-based IPC only) in LC3.

%%%%% Fig. MOOP %%%%%%
%%%%% end %%%%%%%%%%%
In nominal healthy conditions, the proposed architecture has the same performance as a non-switched SPRC-based IPC, which is designed to alleviate significantly blade loads in comparison to the turbine baseline controller.
After the fault is successfully detected and isolated (Fig. \ref{Pic_FD}), the values of the SPRC parameters are switched to the pre-tuned values in order to quickly accommodate the fault, as shown in Fig. \ref{Pic_MOOP} and \ref{Pic_angle}. In the same figures it is possible to notice, as well, that the regular SPRC algorithms may lead to even higher blade loads than the baseline controller during faulty conditions (Fig. \ref{Pic_MOOP}(a)).

\begin{figure}
\centering 
\includegraphics[width=1\columnwidth]{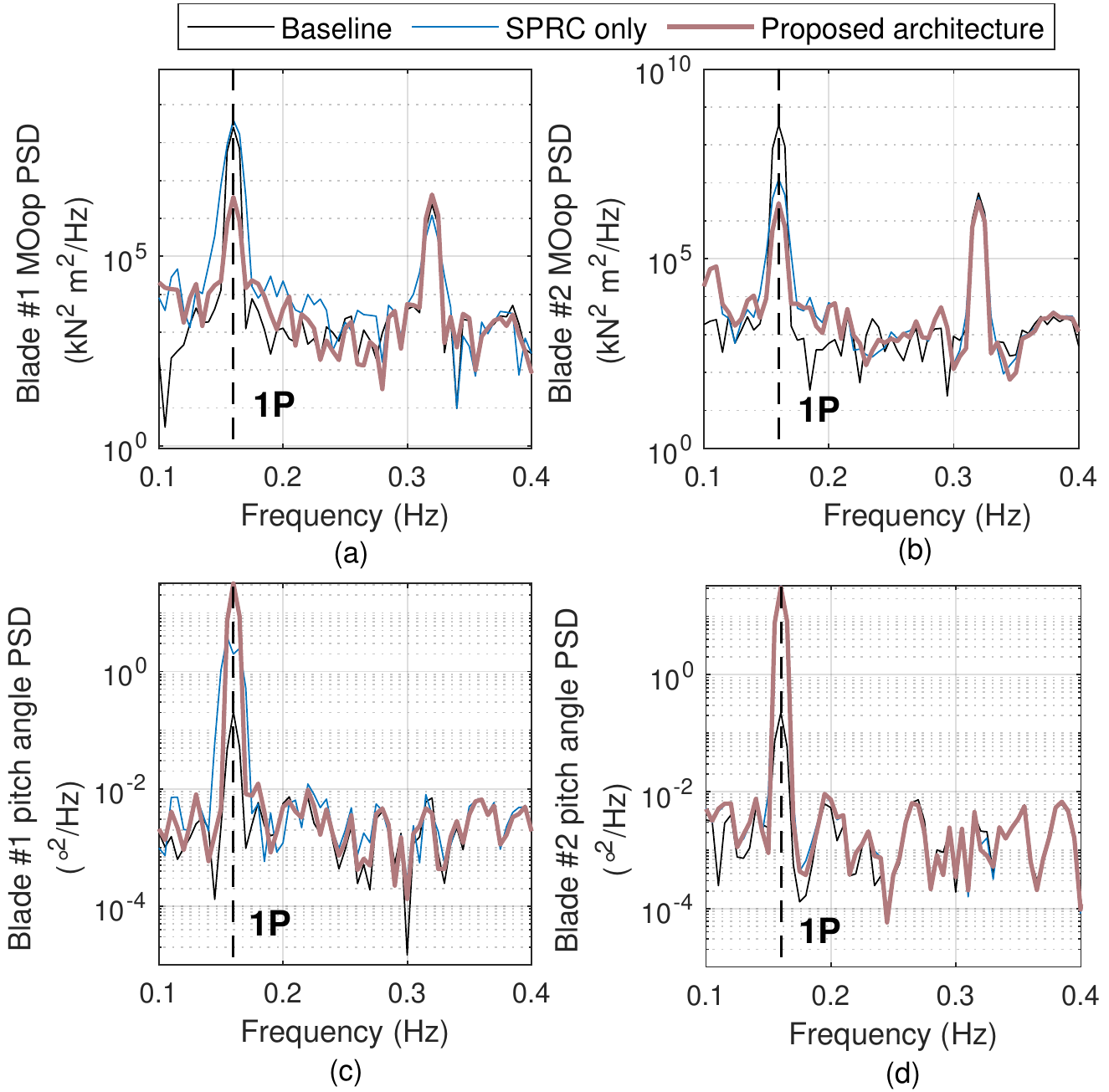}
\vspace{-0.3cm}
\caption{PSD of MOoP and pitch angle. (a) MOoP of blade $\#1$, (b) MOoP of blade $\#2$, (c) Pitch angle of blade $\#1$, (d) Pitch angle of blade $\#2$. Blade $\#3$ is not shown since it is the faulty blade.}
\label{Pic_PSD} %
\end{figure}

\begin{figure}
\centering 
\includegraphics[width=1\columnwidth]{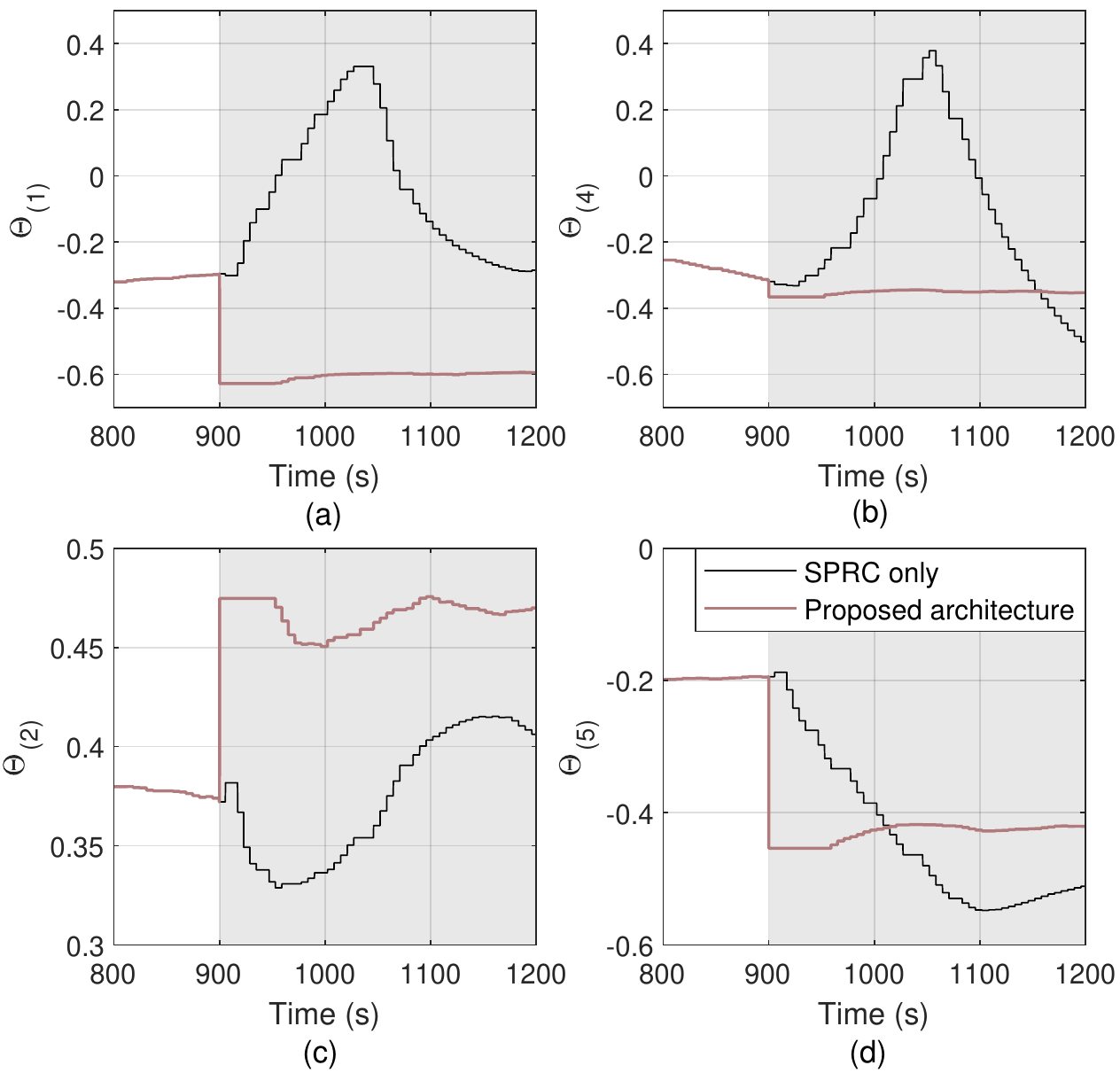}
\vspace{-0.6cm}
\caption{Time series of $\theta$. (a) $\theta_{(1)}$ of blade $\#1$, (b) $\theta_{(4)}$ of blade $\#1$, (c) $\theta_{(2)}$ of blade $\#2$ (d) $\theta_{(5)}$ of blade $\#2$. Time periods of faulty conditions are indicated by a grey background.}
\label{Pic_theta_com} %
\end{figure}

Furthermore, the Power Spectrum Densities (PSDs) of the corresponding MOoP in Fig. \ref{Pic_MOOP} within last 200s simulation periods are illustrated in Fig. \ref{Pic_PSD} for investigations. 
%%%%% Fig. PSD %%%%%
%%%%%%%%%%%%%%%%%%%%%%%%%
It is clear that MOoPs at 1P frequency of blade $\#1$ and $\#2$ are significantly reduced by the pitch angles formulated by the proposed architecture, compared to the baseline controller and SPRC-based IPC only.
%In summary, the comparison demonstrates the effectiveness and benefits of the proposed FTC architecture. 
Particularly, the corresponding $\theta$ in LC3, used to formulate the pitch control demands, are presented in Fig. \ref{Pic_theta_com}.
%As introduced in equation (\ref{eq:control input}) in section \ref{sec:3}, $\theta$ is persistently estimated to formulate the desired pitch control demands. In this case study, $\theta_{(1)}$ and $\theta_{(4)}$ are used for blade $\#$1 while $\theta_2$ and $\theta_{(5)}$ are utilized for blade $\#$2 to generate pitch control demands, respectively.
It is found that the parameter $\theta$, when using the proposed architecture, tends to a value very quickly, which implies a much faster adaptation compared to the SPRC-based IPC only.
%In order to quantify the effectiveness of the proposed architecture, the variance of MOoP and $\theta$, i.e. var($MOoP_{(l)}$) and var($\theta_{(j)}$) where $l=1,2$ and $k=1,4,2,5$, is calculated. Next, the reduction of var($MOoP_l$) and var($\theta_{(j)}$) in all three LCs are listed in Table \ref{table:indicator} to show the benefits of the proposed architecture in the present study.
Similar results are observed in other LCs. %All these results are summarized in Tab. \ref{table:indicator}. 
In detail, the load reductions compared to the baseline controller
%and the reductions of $\theta$ variations compared to the conventional SPRC, 
are calculated and summarized in Tab. \ref{table:indicator}. It is clear from this table  that the MOoP are significantly reduced by the propose architecture. Particularly, the cumulative MOoP are reduced by the proposed combined FD and SPRC-based IPC by $\sim66\%$ on average. In comparison, the scheme based on only the SPRC-based IPC attains an average reduction of $\sim34\%$.

\begin{table}
\setlength{\tabcolsep}{2.3mm}{
\caption{Load reduction in faulty conditions in all LCs*. \label{table:indicator}}
\begin{tabular}{ccccc}
\hline \hline
    & LC1  ($\%$) & LC2 ($\%$)  & LC3 ($\%$)   \\ \hline
%Reduction of OoP bending moment variations \\
\textbf{SPRC-based IPC only} \\
Blade $\#$1 &   41.93    &   60.93       &     13.89          \\
Blade $\#2$ &   6.53    &   55.03          &     70.12           \\
Cumulative  &   23.64    &   56.39          &     21.65              \\
\textbf{Proposed architecture} \\
Blade $\#$1 &  50.53    &    70.06       &        80.36        \\
Blade $\#2$ &  50.54    &   69.00        &       83.28          \\
Cumulative &   50.23    &    67.11      &  80.57
%\\  \hline
%Reduction of theta variations \\
%$\theta_{(1)}$  & 88.82    &      89.07      &    85.55 \\
%$\theta_{(4)}$ & 89.28    &      70.78      &    93.52 \\
%$\theta_{(2)}$  & 89.11    &      85.33      &    5.19 \\
%$\theta_{(5)}$  & 86.84    &      91.73      &    88.07 \\ 
%Cumulative      &          &                  &
\\ \hline \hline
\end{tabular}}
\footnotesize
\vspace{0.1cm}
*The number indicates the reduction of the load variance in $\%$.
\end{table}

%%%%% Fig. theta_com %%%%%
%%%%%%%%%%%%%%%%%%%%%%%%%
% It is clear from Tab. \ref{table:indicator} that the MOoP are significantly reduced by the propose architecture. 
% Particularly, the cumulative MOoP are reduced by the proposed architecture by $\sim66\%$ in all LCs, which is higher than the SPRC-based IPC only by $\sim95\%$. Such a consequence implies the effectiveness of PAS-induced load reduction as well as the benefits in shortening the adaption time.

%In summary, the proposed FTC architecture is able to achieve a fast adaptive fault accommodation for PAS type of fault in FOWTs, which avoiding further damage to other components and allowing continued power generation.
\vspace{-0.15cm}

%% file: sections/5_conclusions.tex
%Implementation of fault accommodation for PAS type of faults in FOWTs has aroused widespread attention recently, in order to reduce cost of energy in wind industry. 
In this paper, a novel architecture is proposed to accommodate the PAS type of faults in a fast adaptive way.  
In detail, a model-based FD scheme is developed for pitch actuators in FOWTs to detect and isolate PAS type of pitch actuator faults. Based on the fault isolation results, a pre-tuned SPRC-based IPC is switched online to accommodate the detected PAS type of faults, whose initial values of the parameters has been tuned offline to match the specific faulty condition.
The effectiveness and benefits of the proposed architecture are illustrated via a case study of a 10MW FOWT in different LCs. 
Results show that the proposed architecture, integrating the FD scheme and SPRC-based IPC, is able to significantly alleviate the PAS-induced loads.
More importantly, the time needed to reduce the PAS-induced loads are significant shortened, which, to some extent, avoids further damage to other components of FOWTs due to the possible improper control demands formulated by the SPRC-based IPC only during the slow adaption time, and thereby allow continued power generation. 
%Future work will include the capability to reduce aerodynamic loads in higher frequency (2P, 3P, etc.). Furthermore, a more general fault tolerant control for severe and non-severe faults in FOWTs will be developed. 
\vspace{-0.7cm}